# Activatability for simulation tractability of NP problems: Application to Ecology

Patrick Coquillard[a], Alexandre Muzy[b], Eric Wajnberg[a]

a: Ecologie moléculaire et comportementale, UMR CNRS-INRA-Université de Nice, 400 route de Chappes, P.O. box 167, 06903, Sophia-Antipolis, France.

b: UMR CNRS LISA Università di Corsica - Pasquale Paoli UFR Drittu, Scenzi suciali, ecunòmichi è di gestioni 22, av. Jean Nicoli, BP 52 20250 Corti, France.

**Summary**

Dynamics of biological-ecological systems is strongly depending on spatial dimensions. Most of powerful simulators in ecology take into account for system spatiality thus embedding stochastic processes. Due to the difficulty of researching particular trajectories, biologists and computer scientists aim at predicting the most probable trajectories of systems under study. Doing that, they considerably reduce computation times. However, because of the largeness of space, the execution time remains usually polynomial in time. In order to reduce execution times we propose an activatability-based search cycle through the process space. This cycle eliminates the redundant processes on a statistical basis (Generalized Linear Model), and converges to the minimal number of processes required to match simulation objectives.

# Introduction

Most of –if not all–biological and ecological systems are strongly influenced by spatial dimensions. Indeed, it is well established that, whatever the particular space scale the systems are considered, the analysis of interactions between organisms, or between organisms and physico-chemical components, is crucial to understand system behaviour and structure. Furthermore, such interactions may occur both in various ways (e.g., secretion of chemical compounds, contacts between individuals, competition for resource, gain or loss of matter and energy, etc.) and at various distances (i.e., from immediate neighbouring to long distances). Finally, such interactions may occur in continuous or discrete modes in both space and time.

Powerful simulators in ecology actually take into account the spatial dimension of interactions (Wu J. and David J. L., 2002; Ratzé et al., 2007). Several techniques are used, some of them allowing to involve both space and time, at a low level of details (e.g., Kendall process, stepping stone models, compartment models, etc.). Among these modelling techniques, the most powerful simulators in ecology belong to the class of "individual based models" (IBM hereafter; also denoted individually oriented models (IOM), Fishwick et al. 1998) which allow integrating spatial interactions at a high level of details. The IBM approach completes the set of usual formal mathematical methods (Grimm 1994, Sultangazin 2004). For instance, differential equations or partial differential equations are very efficient to give a coarse estimation of the evolution of large areas. However, (partial) differential equations are limited for simulating actual biological processes (Grimm and Uchmanski, 1994), particularly when the questions to be address require many details. Ecological modelling often has to account simultaneously for: (1) the diversity of individuals, (2) the spatial heterogeneity of the environment, (3) the changing interaction network (and changes of biotic structures), (4) the discrete and distant interactions between individuals, (5) the random processes and behaviours (i.e., random spatial interactions or movements), etc.

IBMs are often implemented by object-oriented models (Coquillard and Hill, 1997) or by multi-agent models when there is a need to represent an autonomous social behaviour of individuals heading a common goal (Ferber, 1999). In this case, IBMs are usually called agents. In addition, such modelling approach has the main following advantages (Hill & Coquillard, 2007): (1) it allows theoretically the simulation of ecosystems with large sets of species harbouring different behaviours. Moreover, object classes can account for a part of mathematical modelling in order to obtain combined simulations if needed (mixing the discrete and continuous approach). (2) However, a lot of fieldwork always remains necessary as well as a deep knowledge of the modelled species. (3) It takes into account the spatial features of ecosystems that is difficult with partial differential equations (e.g., compartment models), or with the classical Markovian analysis. (4) It provides the possibility to manage, for each individual, the set of parameters the biologist decides to integrate in the model. The management of individuals, and correlatively of their physiological variations, enables model refinement to approach reality according to the detailed level wished by the user.

In a first part, we will show, through a simple example, that (1) introducing spatial relationships between individuals is a prerequisite to maintain a sufficient level of diversity and (2) that such operation requires a probabilistic approach. In a second part we will propose a statistically driven method for reducing the state of space for a model.

## I. Reducing activity

*Example of spatialization necessity*

Let us have a look to the prisoner's dilemma which is the most emblematic problem in the game theory. The problem was initially formalized by W. Tucker. In its classical form, the dilemma is expressed as follows:

*Two suspects are arrested by the police. The police having separated both prisoners visit each of them to offer the same deal. If one testifies for the prosecution against the other and the other remains silent (cooperates), the betrayer goes free and the silent accomplice receives a full 10-years sentence. If both remain silent, both prisoners are sentenced to only six months in jail. If each betrays the other, each receives a five-year sentence. Each prisoner must choose to betray the other or to remain silent. Each one is assured that the other would not know about the betrayal before the end of the investigation. How optimally should the prisoners act?*

In this game, the only concern of each prisoner is to maximize his payoff. Consequently, all rational players should play "testify" (fig. 1) and cooperating is strictly dominated by defecting. As a consequence of such strategy, the game leads to the disappearance of cooperators. But many examples of coexistence of cooperation and selfish behaviours can be found in animal societies and economical situations. How is it possible? M. Nowak and R. May demonstrated in 1992-1995 that introducing spatial dimension in the dilemma, even in an elementary - and somewhat opened to criticism - form, makes such situation possible. They first reformulated the dilemma by introducing a sentence variable $b$ ($b>1$). Then, they distributed players on a grid in which each player has a probability to become a cooperator. This probability is function of states and gains of its immediate neighbours (see fig. 2 for details). As a result of such a transformation, they obtained for some couples *(m, b)* the coexistence of both strategies (see fig. 3). By doing that, however, they brought into the model some probabilistic compounds. Actually, this example illustrates clearly the usual way simulators reproduce spatial interactions.

**If the other prisoner testifies:**
    If I remain silent, I will receive the full 10-years sentence;
    But if I testify, I will only receive a 5-years sentence
**If he does not:**
    If I remain silent, I will receive a 6-monthes sentence;
    But if I testify, I will be free.
*«Whatever his choice, I have interest to testify»*

|         | Silent       | Testify   |
|---------|--------------|-----------|
| Silent  | (-0.5, -0.5) | (-10, 0)  |
| Testify | (0, -10)     | (-5, -5)  |

↓

|         | Silent | Testify |
|---------|--------|---------|
| Silent  | (1 , 1)| (b , 0) |
| Testify | (0 , b)| (0 , 0) |

Figure 1. The original prisoner's dilemma (upper table) is reformulated by introducing *b>1* as a sentence variable (lower table).

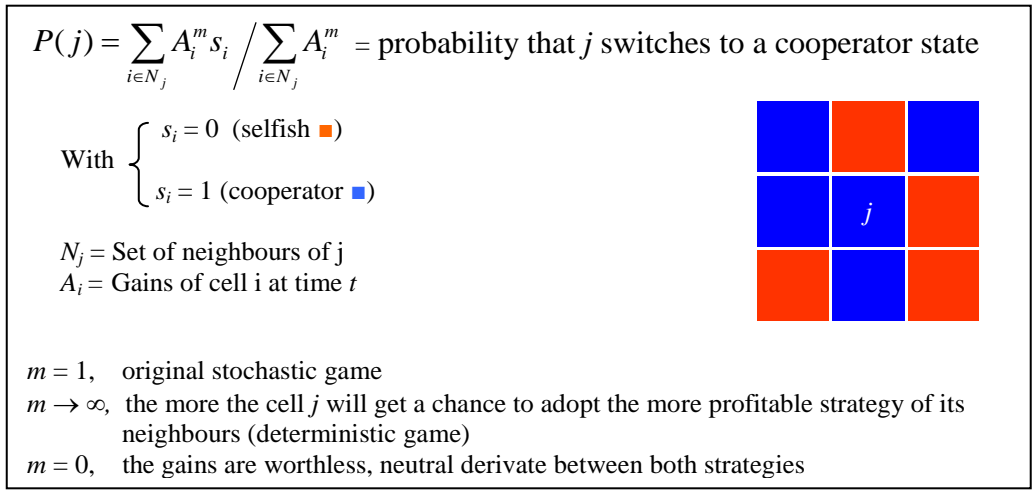

$$P(j) = \sum_{i \in N_j} A_i^m s_i \Big/ \sum_{i \in N_j} A_i^m = \text{probability that } j \text{ switches to a cooperator state}$$

With $\begin{cases} s_i = 0 \text{ (selfish ▪)} \\ s_i = 1 \text{ (cooperator ▪)} \end{cases}$

$N_j$ = Set of neighbours of j
$A_i$ = Gains of cell i at time *t*

$m = 1$, original stochastic game
$m \to \infty$, the more the cell *j* will get a chance to adopt the more profitable strategy of its neighbours (deterministic game)
$m = 0$, the gains are worthless, neutral derivate between both strategies

Figure 2. The spatialized dilemma (after M. Nowak and R. May). *P(j)* is the probability the prisoner *j* has to become a cooperator.

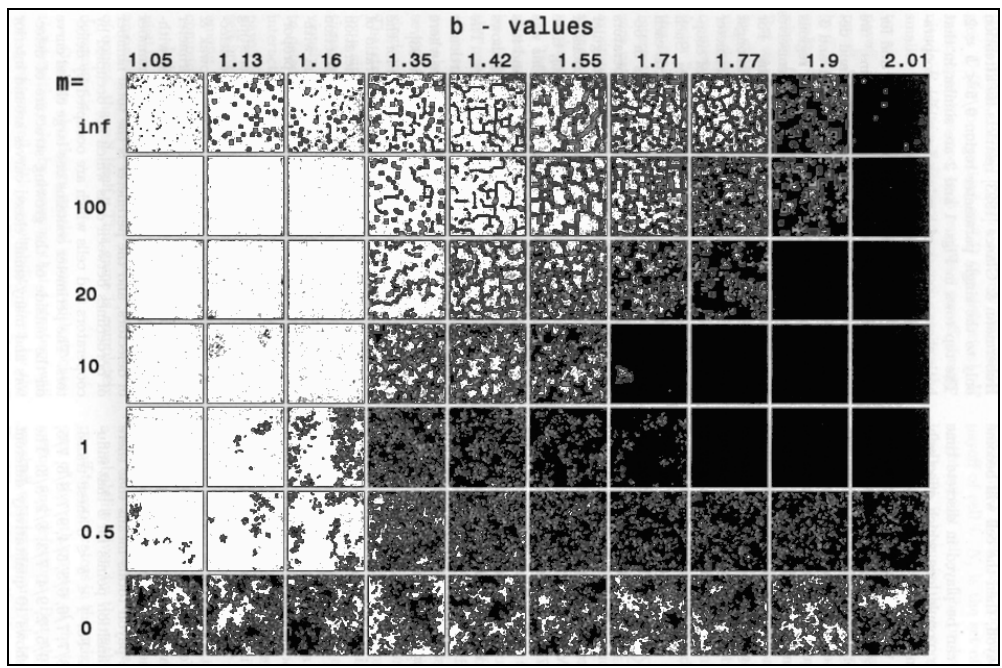

Figure 3. The spatialized prisoner's dilemma in the *(m,b)* plane. Each cell of the grid represents a game of 100 × 100 players. Large areas present the coexistence of cooperators (white) or selfish players (black). In gray colour, the players which have just changed of state. After M. Nowak and R. May (modified).

Obviously, integration of space into simulators leads to a better representation of reality. However, the level of details increases the number of parameters, computations and interactions between parameters. This results in an explosion of the state space and to the intractability of simulations. Therefore, solutions have to be found to reduce the state space and thus enhance tractability.

*Space and NP problems*

According to the previous example, we can conclude that embedding spatial interactions into simulators (designed to model ecological systems) requires, in most cases, a stochastic approach. In most cases, several processes are acting simultaneously in the course of runs to simulate spatial interactions (e.g., various types of competition, seed dispersal, migration, gene flows, chromosome shuffling, chromosomal crossovers, etc.) Considering that each of these processes act on several levels (that can be large), the number of possible trajectories of the system is an exponential function of the number *n* of processes varying on *p* levels, and the time required to find a particular trajectory of the system is $O(n^p)$. In other words, this problem belongs to the NP complexity class problem (see fig. 4). This is the reason why ecologists have early decided to reformulate the statement: "finding a particular trajectory" into "finding the most probable trajectory".

Let us define *activity* of a system as *its number of transitions* and *activatability* as *the probability of transition activation*. If we consider *p* processes varying on [1,...,*n*] levels and that each level can be activated with a probability following a law π, the activatability of the process *i* is (fig. 4):

$$\Delta_i[\pi_i(s_{i1},\ldots, s_{in})],$$

The activity can be estimated as proportional to the number of replicates (*R*), the confidence interval (%) and the number of processes (*p*):

$$A\alpha(R,\%,p)^1$$

The number of replicates *R* depends on σ (the standard error of the response).

Reformulating the original question in "finding the most probable trajectory", we considerably reduced the computation time and escape from the NP-problem trap along with its heuristic solutions (the problem is now solvable in a polynomial time $O(klog(p))$ with *p* processes). However, the problem of activating many stochastic processes is still relevant and some algorithms can be time consuming (for instance, the algorithm AKS which tests the primarity of a number is $O(log(n)^{10.5})$).

These considerations lead to the following question: "How to reduce the number of processes?" Kleijnen and Groenendaal (1992) proposed that building of $2^{(n-1)}$ experimental designs in which each process is either active or inactive, give the same information than a $2^n$ protocol (i.e., involving a half of process combinations). Thus, it is possible to test the effects of each process on the results eliminating redundant ones. The use of $2^{(n-k)}$ protocols is also possible but results in introducing confusion between some interactions and a confusion of the main effects with their interactions.

---

[1] The confidence interval of a mean is calculated as: $\bar{x} - T_{\alpha/2;d}\frac{s}{\sqrt{R}} \leq \mu \leq \bar{x} + T_{\alpha/2;d}\frac{s}{\sqrt{R}}$, where $T_{\alpha/2}$ is given by the student law with $\alpha \leq 0.05$ and *d* is the degree of freedom (df).

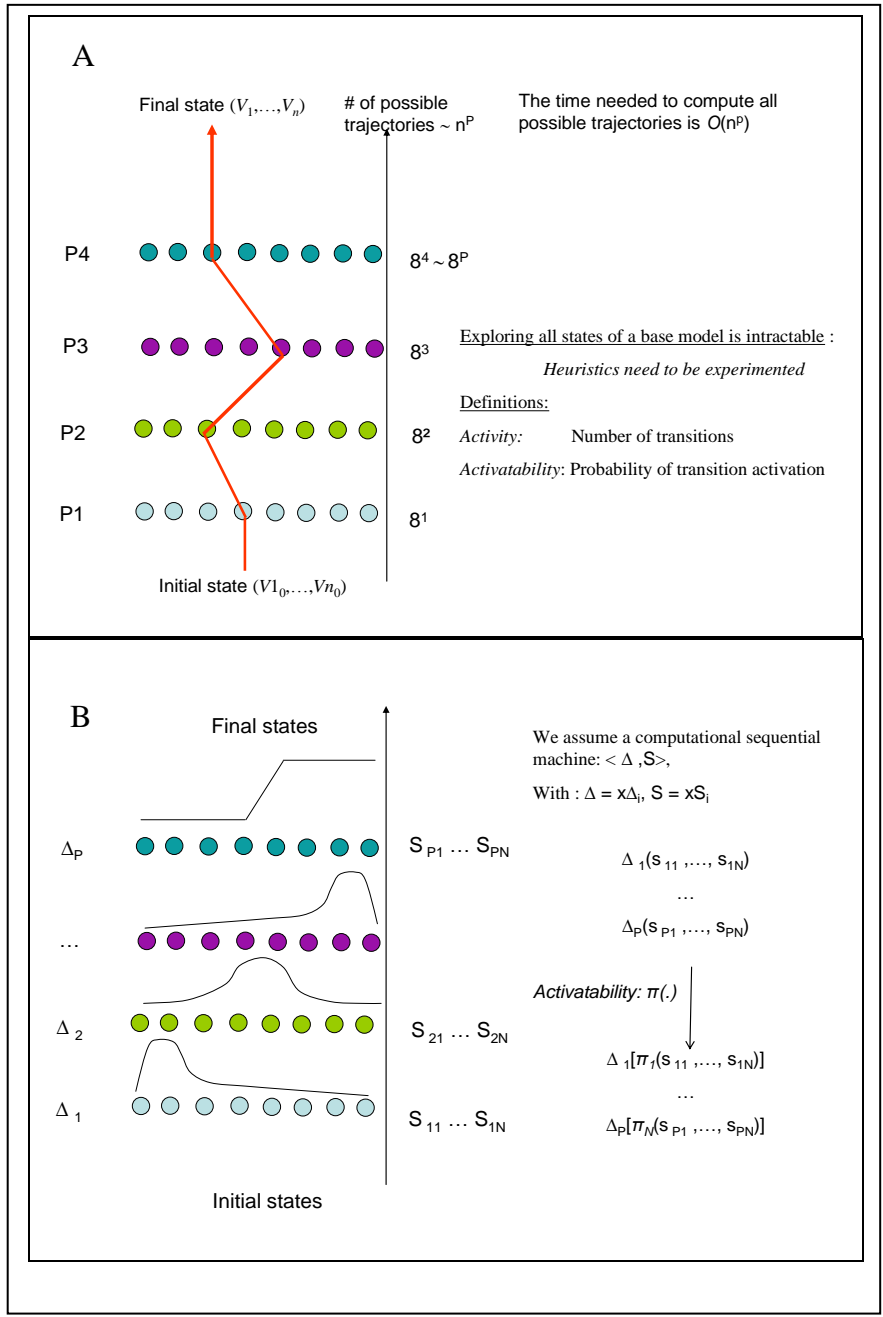

Figure 4. Activity, activatability and processes.

*Proposal of an activatability cycle*

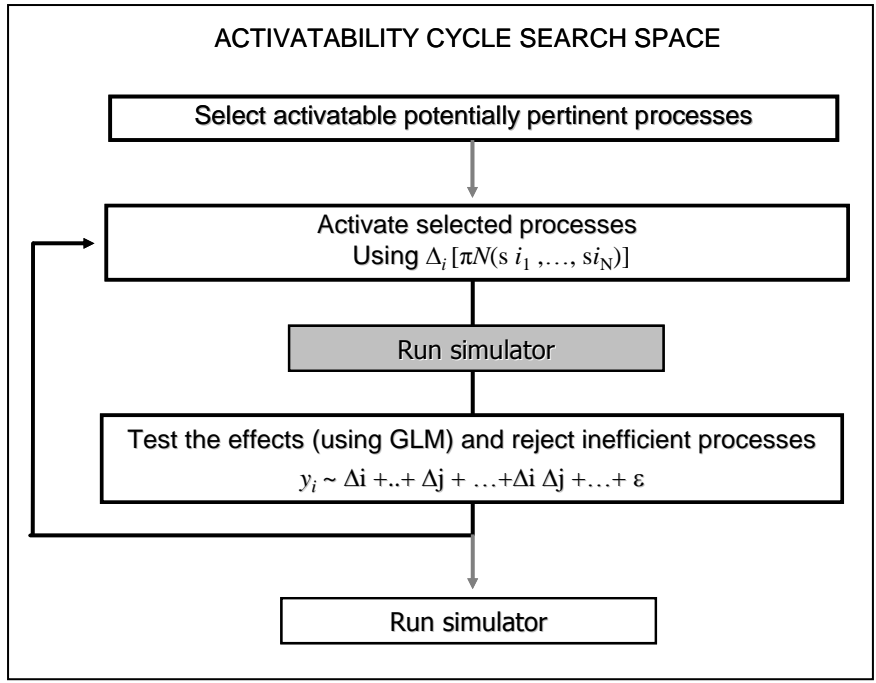

Figure 5. The activatability cycle. Every loop, processes with no significant effects are eliminated.

In this section we will propose a method to build the most parsimonious model from a list of processes. Based on statistical evidences, the method is automatable and allows to embed into the simulator all the process which have a significant effect on the results of simulation, even those on which we have no a priori ideas about their pertinence in the model. However, the method cannot be seen as a validation process. In addition, the proposal must not be confused with the works of Hoffman (2005), who proposed an extended genetic algorithm based method "to accomplish simultaneously parameter fitting and parsimonious model selection" among a list of candidate models. The proposed cycle of activatability (fig. 5) is based on a complete design protocol ($2^n$ protocol). At every cycle, main effects and their interactions on the response $y$, can be tested through a Generalized Linear Model (Nelder and Wedderburn, 1972).

$$E(y_i) = \beta_0 + \beta_1 \Delta_1 + \cdots + \beta_i \Delta_i + \cdots + \beta_k \Delta_i . \Delta_j + \cdots + \varepsilon \qquad (1)$$

where $y_i$ is the dependant variable[2], $\Delta_i$ ($i \in [1..n]$) the principal effects (or independent variables), $\Delta_i . \Delta_j$ the interaction between the effects (sometimes called "product terms") and $\varepsilon$ is a random error. Equation (1) is thus a linear regression. Quadratic effects can also be included in the regression (i.e $(\Delta_i)^2$)). In a GLM, it is assumed that $\varepsilon$ obeys to one function of the exponential family (Normal, Poisson, Binomial, etc.). The $\beta_k$ parameters represent the variation of $E(y_i)$ when the $k^{th}$ variable move of one unit, the remaining variables being unchanged. Formally:

---

[2] In addition, the dependant variable $y$ can be transformed by means of a link function. This is usually the case when the $y_i$ responses do not follow the Normal distribution. Note that GLM assumes that the $y_i$ observations are independent.

$$\beta_i = \frac{\partial E(y_i)}{\partial \Delta_i}$$

Equation (1) is solved by the usual matrix method for multiple regressions. In the general case, the resulting model is then tested against the $y_i$ responses by means of an analysis of variance (ANOVA) which leads to: $R^2 = \frac{\sum(\hat{y}_i - \bar{y})^2}{\sum(y_i - \bar{y})^2}$, called the correlation ratio, where $\hat{y}_i$ is the predicted response, $\bar{y}$ the average response and $y_i$ the observed response. $R^2$ gives the amount of variation of the $y_i$ which is explained by the model.

In addition, it can be demonstrated that $F = \frac{R^2/k}{(1-R^2)/(n-k-1)}$ follows a Fisher distribution with $k$ and $k$-$n$-1 degrees of freedom. In such conditions, we can reject the null hypothesis $H_0$: $\beta_1 = \beta_2 = \ldots = \beta_k = 0$, if $P(F \geq F_{calculated}) < \alpha$, with $\alpha = 0.05$. However, even if we reject the null hypothesis, this does not imply that all variables of the model have a significant contribution to the response $y$. To decide if a particular variable $j$ has a significant contribution to the response we calculate $F = \frac{\beta_j^2}{s^2(\beta_j)}$ and reject the hypothesis $H_0$: $\beta_j = 0$, if $F > F_{\alpha;1,n-k-1}$. To test successively each variable of the model, a stepwise[3] procedure eliminates and introduces the variables in the model of the response $y$ (usually the *p*-value *<0.05* criterion is used).

Such a procedure takes advantages from allowing to embed into the simulator all the variables (= processes) the user wants to test – with no *a priori* exclusion. Another advantage is that processes can be aggregated into sets which can be treated as active units. In this case, the user attempts to measure the effects of some global activities (sexual reproduction for instance) on a response (population dynamics involving both sexual and asexual reproduction of plants).

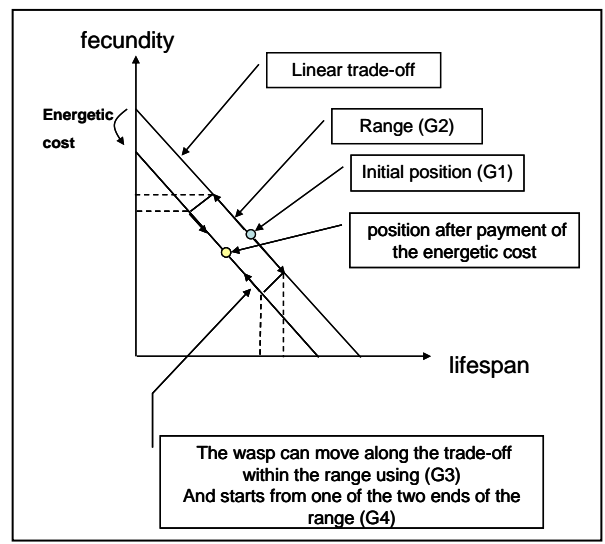

Figure 6. The life span-fecundity trade-off of a parasitoïd wasp.

---

[3] Backward and forward methods are also used. Each of them introduces or eliminates step by step the variables into the model.

*Application*

In this application, four processes are represented by four real values coded on *16 bit*-structure called genes (see figure 6). The animal (a parasitoid[4] wasp) has an initial position along a trade-off (coded by gene *G1*) and can change its reproductive strategy throughout its life thanks to gene *G2*, which defines a range. The wider *G2*, the heavier the cost to pay (in term of energy). The cost reduces both fecundity and lifespan of the animal. The wasp can move within the range thanks to gene *G3*, which is a parameter of a Bayesian estimator re-evaluated each ten time steps of its life as follows:

```
Posterior = prior × G3 + (1-G3) × posterior;
Fecundity = Fecundity + G2 × posterior;
```

When a wasp encounters a patch of hosts, the number of hosts it attacks obeys to a saturation function. Thus, its velocity of attacking hosts decreases, following an exponential function. When its velocity has reached the average velocity calculated on the basis of the average environment richness, it leaves the patch and tries to find a new one. The cycle "foraging for hosts on patches and travelling between patches" is repeated until the wasp has reached the end of its life or has exhausted its potential fecundity.

The four genes are encapsulated into a single chromosome. Each wasp holds a single chromosome. The goal of the simulation consist in finding the vector {*G1,…,G4*} which maximises the score of the wasp, *i.e.* the number of eggs laid throughout its life. The score maximization is obtained by means of a genetic algorithm (GaLib, MIT, 1997-2007).

Basically, the four genes {*G1,…,G4*} are variables. However, each of them induces the call of several functions in the code and modifies the behaviour of the wasp. For instance, the gene G2 can strongly modify the phenotypic plasticity of the animal (i.e. its ability to adapt its fecundity/lifespan ratio to the environment characteristics) and influence its score. That is why we will now consider the {*G1,…,G4*} genes as processes instead of variables.

First cycle.
In this application, 2625 experiments (10 replicates each) were done. The results showed that there was no significant effect of *G4* on scores, whatever the initial conditions in which wasps had to evolve. Consequently, the process directed by *G4* was dropped.

Second cycle.
Significant effects on {*G1, G2, G3*} were found, and the three processes clearly acted on animal scores (table I).

---

[4] An organism that lives at the expense of another (its host), impedes its growth and eventually kills it. Insect parasitoids, which are often very tiny, attack a single organism (plant or animal), from which they derive everything they need for their own growth and reproduction. One way a parasitoid does this is by laying its eggs in the body of the host insect (from Natural Canadian Research document).

| | G1 | | G2 | | G3 | |
|---|---|---|---|---|---|---|
| Source of variation | F Value (calculated) | Pr > F | F Value (calculated) | Pr > F | F Value (calculated) | Pr > F |
| stability of environment (1) | 263.81 | **<.0001** | 24.17 | **<.0001** | 2.04 | 0.0864 |
| inter patch travel time (2) | 222.31 | **<.0001** | 91.23 | **<.0001** | 3.22 | **0.0121** |
| (1).(2) | 7.97 | **<.0001** | 1.78 | 0.028 | 0.74 | 0.7534 |
| energetic cost (3) | 2.34 | 0.0532 | 56.46 | **<.0001** | 4.15 | **0.0023** |
| (1).(3) | 1.11 | 0.3409 | 2.96 | **<.0001** | 1.29 | 0.1968 |
| (2).(3) | 1.28 | 0.2031 | 4.85 | **<.0001** | 0.59 | 0.8926 |
| averaged # hosts on patchs (4) | 3045.64 | **<.0001** | 25.14 | **<.0001** | 10.95 | **<.0001** |
| (1).(4) | 20.85 | **<.0001** | 11.48 | **<.0001** | 1.19 | 0.2392 |
| (2).(4) | 56.15 | **<.0001** | 11.05 | **<.0001** | 2.05 | **0.0019** |
| (3).(4) | 1.25 | 0.1834 | 1.45 | 0.0741 | 1.26 | 0.1794 |
| stochasticity (5) | 24.37 | **<.0001** | 2.19 | 0.1126 | 1.66 | 0.1899 |
| (1).(5) | 4.27 | **<.0001** | 1.08 | 0.3755 | 1.14 | 0.3346 |
| (2).(5) | 1.47 | 0.1612 | 1.9 | 0.0553 | 0.52 | 0.844 |
| (3).(5) | 0.36 | 0.94 | 0.49 | 0.8676 | 1.38 | 0.202 |
| (4).(5) | 6.3 | **<.0001** | 3.07 | **0.0003** | 1.08 | 0.375 |

Table I. ANOVA test on the model obtained by the GLM procedure. Sources (factor) of variation (Δ*i*) and interactions (Δ*i*.Δ*j*) of the linear model are indicated in the left column. Effects on {*G1,…,G4*}: *F* values and their probability to be greater than the theoretical values of the Fisher law are indicated for {*G1,…,G3*}; values for *G4* are omitted since probabilities were systematically ≥ 0.05. Significant effects are indicated in bold.

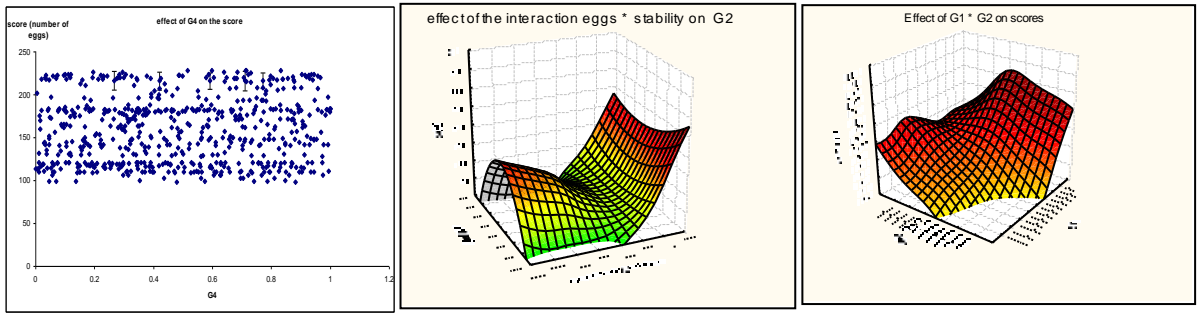

Figure 7. Simulation results (population size = *500*, number of generation = *300*, number of replicates = *50*). Left: no effect of *G4* on scores was detected according to the ANOVA results. Centre and right: after removal of *G4*, significant effects of initial conditions on *{G1, G2, G3}* and of *{G1, G2, G3}* on scores were identified.

| | With GA4 | Without GA4 |
|---|---|---|
| Code size (compiled) | 842 757 | 842 629 |
| # functions | 58 | 52 |
| # max of calls | 1 992 190 | 1 234 692 |
| Virtual mem. (RES + swap) | 3 380 000 | 3 380 000 |
| RES | 1 572 000 | 1 572 000 |
| Execution time | 8mn47.661 | 6mn59.502 |

Table II. Comparison between simulations embedding process *G4* or not (50 replications of a run initialized with a single combination of parameters).

Clearly, results (table I & fig. 7) showed that process *G4* was redundant. Removing *G4* makes the simulations faster than simulations embedding the four processes, and gain is about *20%*. Most execution time reduction was a consequence of the reduction of function calls (*1 234 692* versus *1 992 190*). On the contrary, code size and memory size remained unchanged (table II).

However, it is clear that if the activatability cycle was designed to select processes contributing significantly to the response of the simulator, it does not constitute a validation of the resulting model. Indeed, the validation phase must be engaged after the selection of processes has been achieved, as it is usually done in a classical approach. Lastly, if the resulting model can be validated by comparison with experimental data corresponding to the protocol design, the model has few chances to be validated when confronted to other data. In this case, the experimental design must be rebuilt and the activatability cycle reengaged.

## II. Monitoring the activity of a simulated system

The table I showed that the activatability cycle we used resulted in a substantial reduction of activity of the simulator. However, the activity itself was indirectly estimated through both computing time and number of functions called over the runs. Consequently, the monitoring of activity throughout time remains an opened question.

In the information theory, the entropy is considered as a measure of the disorder of the system. Let us consider four simple binary units {*G1,...,G4*} which can be in one of the two states: *Gi* = 0 or *Gi* = 1. We thus have 16 possible states:

| G1 | 1 | 1 | 1 | 1 | 0 | 0 | 0 | 0 | 0 | 0 | 0 | 0 | 1 | 1 | 1 | 1 |
| G2 | 1 | 1 | 1 | 1 | 1 | 1 | 1 | 1 | 0 | 0 | 0 | 0 | 0 | 0 | 0 | 0 |
| G3 | 1 | 1 | 0 | 0 | 1 | 1 | 0 | 0 | 1 | 1 | 0 | 0 | 1 | 1 | 0 | 0 |
| G4 | 1 | 0 | 0 | 1 | 1 | 0 | 1 | 0 | 1 | 0 | 1 | 0 | 1 | 0 | 1 | 0 |

If the states have equal probabilities, the probability of each state is: $(1/(\text{number of states}))2^{-N}$, where *N* is the length of a state. Thus, we find the entropy of each unit: $s = \frac{1}{N}\log_2(2^N) = 1$. Consequently, if the 16 states are not of equal probability we find that the entropy *per unit* is smaller than 1. The entropy of the system can be computed as:

$$S = -\sum p_i \log_2(p_i)$$, where $p_i$ is the probability of each state.

In such conditions, we can ascertain that the entropy $S_t$ of the system at the instant *t* is limited to the range: $S_{\min}(G) < S_t(G) < S_{\max}(G)$. Indeed, $S_{min}$ represents the particular case in which there is only one possible state (i.e. one possible combination of {*G1,...,G4*}) and $S_{max}$ represents the case where all possible combinations have equal probabilities ($S_{max} = 4$ in this small example).

In a general case, the processes {*G1,...,G4*} can take several values as we saw in the application section. Because of the stochasticity of the simulation, one usually conducts simulations through replicates in order to obtain averaged values $\{\overline{G1},...,\overline{G4}\}$ and associated variances *at each time step of the runs*. It can be reasonability admitted that the probability density function of *G* conforms to a multidimensional Gaussian (in a majority of cases at least). A measure of the entropy can be obtained by establishing the variance-covariance symmetric matrix Σ ($N \times N$) of the vector $\{\overline{G1},...,\overline{G4}\}$ and next by the calculation of its determinant[5]. This value gives the total amount of information diminished by the interactions between the processes. It is also an approximation of the number of possible combinations at each step of time, i.e. a hyper volume of the dispersal of the state space. This hyper volume corresponds to the activatability state space of the system. That is, at every time step, the states of the system can change. Considering a state change of the system as an activity (i.e., constraining the definition of the *activity* term) of the system, its activatability corresponds to its possible state changes or activities. Under these assumptions, the entropy $S_t(Gi)$, at the *t* instant, of the process *Gi* is given by (Ahmed and Gokhale, 1989) : $S_t(G_i) = (1/2)\ln(2\pi e \sigma_i^2)$ where $\sigma_i^2$ is the *i*th element of the diagonal of Σ, and the entropy (or differential entropy) $S_t(G)$, is then given by:

$$S_t(G) \leq (1/2)\ln(2\pi e)^N |\Sigma|$$

$S_{min}$ and $S_{max}$ do not constitute some likely/sustainable situations. The former represents the case in which the system is fixed and is unable to adapt its behaviour to a fluctuant environment. The later characterises a system which is highly adaptable (it can face any situation), but with a too high cost of energy (e.g., for natural systems) or in terms of resources and computation time (virtual systems; see also the figure 6)). We can thus ascertain that the environment in which a system evolves imposes some constraints to the system so that it must adopt a certain level of disorder (positive entropy). This disorder allows the system to face the fluctuations of the environment – within a fixed range – to the extent that it pays an energetic price corresponding to such flexibility.

**Conclusion**

We conclude this paper in enumerating the following traits of the proposal in two short sections:

Advantages
- The proposal of the activatability cycle is automatable.
- The method allows embedding into the simulator all the processes wished by the user with no *a priori* exclusion.
- Processes are activated according to their statistical effects on results.

Drawbacks and troubles
- Reformulating the question, one loses the prediction (statistical results) but recognizes the dimension of complexity in the scientific explanation.
- The necessity of replicates strongly diminishes the information about spatial results. Thus, spatial trajectories of particular interest cannot be identified.
- The selection of processes must be a conservative operation (the internal coherence must be preserved).

---

[5] sometimes called generalized variance

**Acknowledgment**

We would like to warmly thank Pr. Thomas S. Hoffmeister who participated to the elaboration of the conceptual model of the fecundity-lifespan tradeoff of the parasitoïd wasp we used as an example of application.